\RequirePackage{fix-cm}

\documentclass[smallextended]{svjour3}       

\usepackage{color}
\usepackage{amssymb}
\usepackage{graphicx}
\usepackage{epstopdf}
\usepackage{subfigure}
\usepackage{algorithm}
\usepackage{amsmath}
\usepackage{bm}
\usepackage{braket}
\usepackage {lineno}

\journalname{Quantum Information Processing}

\begin{document}
\bibliographystyle{unsrt}

\title{Performance of Grover's search algorithm with diagonalizable collective noises}

\author{Minghua Pan \and  Taiping Xiong \and Shenggen Zheng\footnote{Corresponding author}}

\institute{Minghua Pan \at
Guangxi Key Laboratory of Cryptography and Information Security, Guilin University of Electronic Technology, Guilin 541004, China\\
Department of Physics, Tsinghua University, Beijing 100084, China\\
\and Taiping Xiong \at
Guangxi Key Laboratory of Image and Graphic Intelligent Processing, Guilin University of Electronic Technology, Guilin 541004, China\\
\and Shenggen Zheng \at Peng Cheng Laboratory, Shenzhen 518055, China\\
\email{\{panmh,xiongtp\}@guet.edu.cn; zhengshg@pcl.ac.cn.}
  }

\date{Received: date / Accepted: date}

\maketitle

\begin{abstract}

Grover's search algorithm (GSA) is known to experience a loss of its quadratic speedup when exposed to quantum noise. In this study, we partially agree with this result and present our findings.
First, we examine different typical diagonalizable noises acting on the oracles in GSA and find that the success probability decreases and oscillates around $1/2$ as the number of iterations increases. Secondly, our results show that the performance of GSA can be improved by certain types of noise, such as bit flip and bit-phase flip noise. Finally, we determine the noise threshold for bit-phase flip noise to achieve a desired success probability and demonstrate that GSA with bit-phase flip noise still outperforms its classical counterpart.
These results suggest new avenues for research in noisy intermediate-scale quantum (NISQ) computing, such as evaluating the feasibility of quantum algorithms with noise and exploring their applications in machine learning.

\end{abstract}

\section{Introduction}{\label{SecInt}}

In the past few years, quantum computation has seen remarkable progress in both hardware and software. In 2019, the superconducting quantum computing device `Sycamore' made headlines for achieving quantum supremacy \cite{Arute}. The following year, the photonic quantum computing device `Jiuzhang' demonstrated the superiority of quantum computing \cite{Zhong} . In 2022, `Zuchongzhi 2.1' \cite{Zhu2022}  accomplished a sampling task that was roughly 6 times more challenging than what `Sycamore' could handle in classical simulation. These breakthroughs demonstrate the rapid development and increasing capability of quantum computation.

Despite advances in quantum computation, creating a fault-tolerant quantum computer with millions of low-error, long-coherence qubits remains a significant challenge. Currently, we are in the noisy intermediate-scale quantum (NISQ) era, characterized by devices that consist of dozens or hundreds of noisy qubits with limited coherence and imperfect operations \cite{Preskill,Corcoles}. Despite the limitations of NISQ devices, researchers have proposed algorithms that can still achieve quantum advantages in various fields, including physics, machine learning, quantum chemistry, and combinatorial optimization \cite{Cai,Situ,Havlicek,McArdle,Huang,Bharti}. However, critical questions remain about how noise affects the results of existing quantum algorithms and whether quantum advantages can still be achieved within acceptable noise levels. These questions are increasingly important in the NISQ era.

Numerous studies have been conducted in an effort to understand the impact of noise on quantum algorithms in the noisy intermediate-scale quantum (NISQ) era. One of the algorithms that has received a significant amount of attention is Grover's search algorithm (GSA) \cite{Grover97,Nielsen}, as it is known for its quadratic speedup and various applications. Over the years, GSA has been the subject of many research projects, including generalizations and improvements \cite{Long99,Hoyer00,Galindo00,Biham02,Yoder14,Tulsi15,Zhang20}, the analysis of its quadratic acceleration mechanism \cite{Braunstein,Rungta,Cui,Pan17,Pan191,Shi,Pan192}, and experimental verification and implementation \cite{Godfrin17,Chakrabarty17,Srinivasan18,Wang21}. These efforts have resulted in a wealth of knowledge about the algorithm and its potential applications.

Over the years, numerous studies have been conducted to evaluate the performance of GSA in different noisy environments. In 1999, Pablo-Norman \textit{et al.}~\cite{Pablo} analyzed the impact of Gaussian noise on each iteration during the search for a single target. In 2000, Long \textit{et al.}~\cite{Long} discovered that the main quantum gate errors were due to the systematic error of the phase flip gate and the random error of the Hadamard transform. Kapira \textit{et al.}~\cite{Shapira} studied the effects of unitary noise caused by small disturbances and drift, while Shenvi \textit{et al.}~\cite{Shenvi} investigated the robustness of GSA in the presence of a random phase error in the oracle.
Using a perturbative method, Azuma \cite{Azuma} analyzed the effects of phase flip errors on the system. Zhirov \textit{et al.}~\cite{Zhirov} evaluated the impact of dissipative decoherence on GSA using quantum trajectories. The decoherence effect on GSA was also studied by modeling the noise as a depolarizing channel \cite{Salas,Botsinis}. Gawron \textit{et al.}~\cite{Gawron} showed that the quantum search algorithm was more efficient than its classical counterpart under low noise from a computational complexity perspective. In 2018, Rastegin \cite{Rastegin} investigated the degradation of GSA under collective phase flips to the oracles.
In recent years, researchers have found that local noise such as the local depolarized channel \cite{Cohn} and localized dephasing \cite{Reitzner} are much worse than global noise. In 2020, Wang \textit{et al.}~\cite{Wang20} used IBM Qiskit to simulate the GSA under various types of noise and predicted typical gate error bounds.

However, accurately determining the complex and unpredictable real-world noises is challenging. In specific applications, the implementation of GSA will be carried out using specific circuits and gates. In these cases, the impact of various noises, such as global and local noise and isolated gate errors affecting individual physical qubits, can be analyzed.

From the standpoint of algorithm design, GSA is a quantum algorithm that relies on an oracle. In fact, oracle-based algorithms are common in quantum computing and can be seen in various algorithms, including the Deutsch-Jozsa algorithm \cite{Deutsch92}, the Simon algorithm \cite{Simon94}, and other computational complexity algorithms \cite{ambainis2014exact,Buhrman02,Iriyama12,qiu2020revisiting}.
In quantum search algorithms, the number of iterations or queries to the oracle is referred to as the query complexity. General oracle-based algorithms analyze the system's behavior through its input and output rather than its internal structure. As the internal structure of the oracle remains unknown, it can only be treated as a whole. In this case, the noise will also be viewed collectively.
Our research focuses on studying the performance of GSA in the presence of collective noise acting on the entire oracle. To simplify the calculation, we analyze the diagonalization of the noise and incorporate it into our study. Through a combination of theoretical analysis and numerical calculations, we aim to draw universal conclusions.
Taking three types of flip noises as examples of diagonalizable noise, we have made some novel and intriguing findings.

The organization of the paper is outlined as follows: In Section 2, we provide an overview of GSA and several noise models, followed by a presentation of GSA with noise in the Bloch representation. In Section 3, we delve into the concept of diagonalizable noise and examine the effects of diagonalizable noise on GSA in the Bloch representation.
Sections 4 and 5 present our investigation into the performance of GSA in the presence of different types of typical diagonalizable collective noises, through both theoretical analysis and numerical experiments.
In Section 6, we discuss the significance of our work. The conclusion, which includes a summary and suggestions for future research, is given in Section 7.

\section{GSA with noise in Bloch presentation}\label{Sec2}
In this section, we give an overview of GSA and various noise models, followed by the presentation of GSA with noise in the Bloch representation.

\subsection{Grover's search algorithm}\label{GSA}

Suppose we are tasked with searching through a database containing $N=2^n$ items. Instead of searching for the targets directly, our goal is to find the index of one of the $m$ targets that satisfy specific conditions. GSA \cite{Grover97,Nielsen} is a quantum algorithm designed for this purpose.
GSA begins with an equal superposition state of all computational basis states:
$\ket{\psi_0}=H^{\otimes n}\ket{0}^{\otimes n} =\frac{1}{\sqrt{N}}\sum_{x=0}^{N-1}\ket{x}$,
where $H=\frac{1}{\sqrt{2}}\begin{bmatrix} 1 & 1 \\1 & -1 \end{bmatrix}$ is a Hadamard matrix. To make the notation simpler, we denote the superposition of all non-target states as $\ket{\chi_0}=\frac{1}{\sqrt{N-m}}\sum_{x_{n}}\ket{x_{n}}$ and the superposition of all target states as $\ket{\chi_1}=\frac{1}{\sqrt{m}}\sum_{x_s}\ket{x_s}$.

By using $\theta/2=\arcsin\sqrt{m/N}$, the initial state $\ket{\psi_0}$ can be expressed as:

\begin{eqnarray}\label{psi0}
\ket{\psi_0} =\cos\frac{\theta}{2}\ket{\chi_0}+\sin\frac{\theta}{2}\ket{\chi_1}.
\end{eqnarray}

Then, GSA performs a subroutine, known as the \textbf{G iteration}, which includes four basic operations: $G=H^{\otimes n}SH^{\otimes n}O$. This consists of the oracle $O$, which distinguishes between the target and non-target states by functioning as $O\ket{x}=(-1)^{f(x)}\ket{x}$, where $f(x)=1$ if $x$ is the index of a target state, and $f(x)=0$ otherwise. The conditional phase shift operator $S$ also acts as an oracle, differentiating the state $\ket{0}$ from other states by transforming $S\ket{x}=-(-1)^{\delta_{x0}}\ket{x}$. Letting $R=H^{\otimes n}SH^{\otimes n}$, it has been shown that $R=2\ket{\psi_0}\bra{\psi_0}-I$, where $I$ is the identity matrix. In the two-dimensional space spanning ${\ket{\chi_0},\ket{\chi_1}}$, the oracle $O$ performs a reflection about $\ket{\chi_0}$, while $R$ performs a reflection about $\ket{\psi_0}$.

The state after $t$ iterations of the $G$ iteration can be expressed as:
\begin{equation}
\ket{\psi_t} = G^t\ket{\psi_0} = \cos(\theta_t)\ket{\chi_0} + \sin(\theta_t)\ket{\chi_1},
\end{equation}
where $\theta_t = (2t + 1)\frac{\theta}{2}$.
The success probability of finding one of the target states in $\ket{\chi_1}$ is given by $P(t) = \sin^2(\theta_t)$. The optimal number of iterations for GSA is $T = \left\lfloor\frac{\pi}{4}\sqrt{\frac{N}{m}}\right\rfloor$. This is because for $m\ll N$, $T$ iterations can achieve the highest success probability.
It is worth noting that GSA has a quadratic speedup compared to classical algorithms, which would require $\Omega(N)$ iterations to achieve the same result.

\subsection{Noise models in Bloch}

The power set of tools for describing quantum noise and the behavior of open quantum systems is referred to as quantum operations \cite{Nielsen}. A \textbf{quantum operation $\varepsilon$} is a transformation that maps an initial state $\rho$ to a final state $\varepsilon(\rho)$.
Quantum operations can be represented in a convenient form known as the operator-sum representation.

The operator-sum representation is used to model the behavior of real quantum systems, as shown in the following equation:
\begin{eqnarray}\label{Esum}
\varepsilon(\rho)=\sum_k E_k\rho E_k^\dag,
\end{eqnarray}
where $E_k$ are the operation elements and $\sum_k E_k^\dag E_k=I$ if the quantum operation is trace-preserving.
There are several established noise models based on a quantum system's interaction with different environments.
For a single qubit, the most commonly used operations to describe quantum noise models are the depolarizing channel, flip channel, and damping channel.

The flip channel consists of three types of noise: the \emph{bit flip} (BF), the \emph{phase flip} (PF), and the \emph{bit-phase flip} (BPF). The \emph{bit flip channel} flips the state from $\ket{0}$ to $\ket{1}$ or from $\ket{1}$ to $\ket{0}$ with probability $1-p$, resulting in $X$ noise. The \emph{phase flip channel} introduces a phase flip with probability $1-p$ and is denoted by $Z$ noise. The \emph{bit-phase flip channel} is a combination of both the bit flip and phase flip channels and can be expressed as $Y=iXZ$. The Pauli matrices $X$, $Y$, and $Z$ are represented by $X=\begin{bmatrix} 0 & 1 \\ 1 & 0\end{bmatrix}$, $Y=\begin{bmatrix} 0 & -i \\ i & 0 \end{bmatrix}$, and $Z=\begin{bmatrix} 1 & 0 \\ 0 & -1 \end{bmatrix}$, respectively. The operation elements and their impact on the states in the Bloch vector for these three flip channels are shown in Table \ref{Tab1}.

\begin{table}[H]
\caption{Noise Models}\label{Tab1}
\resizebox{0.8\textwidth}{!}{
\begin{tabular}	{|c|c|c|c|}
\hline
Noise &$E_0$ & $E_1$ & $\{r_x,r_y,r_z\}$ after noise acts \\ \hline
BF&$\sqrt{p}\left(\begin{array}{cc}
      1 & 0 \\
      0 & 1 \\
    \end{array} \right)$
& $\sqrt{1-p}\left(\begin{array}{cc}
      0 & 1\\
      1 & 0 \\
    \end{array}\right)$
&$\{r_x,(1-2p)r_y,(1-2p)r_z\}$ \\ \hline
PF&$\sqrt{p}\left(\begin{array}{cc}
       1& 0 \\
      0 & 1 \\
    \end{array} \right)$
& $\sqrt{1-p}\left(\begin{array}{cc}
      1& 0 \\
      0 &-1 \\
    \end{array}\right)$
&$\{(1-2p)r_x,(1-2p)r_y,r_z\}$ \\ \hline
BPF&$\sqrt{p}\left(\begin{array}{cc}
       1& 0 \\
      0 & 1 \\
    \end{array} \right)$
& $\sqrt{1-p}\left(\begin{array}{cc}
      0 & -i \\
      i& 0 \\
    \end{array}\right)$
&$\{(1-2p)r_x,r_y,(1-2p)r_z\}$ \\ \hline
AD&$\left(\begin{array}{cc}
      1 & 0 \\
      0 & \sqrt{\gamma} \\
    \end{array} \right)$
& $\left(\begin{array}{cc}
      0 & \sqrt{1-\gamma} \\
      0 & 0 \\
    \end{array}\right)$
&$\{\sqrt{\gamma}r_x,\sqrt{\gamma}r_y,1-\gamma+\gamma r_z\}$ \\ \hline
PD&$\left(\begin{array}{cc}
      1 & 0 \\
      0 & \sqrt{\gamma} \\
    \end{array} \right)$
& $\left(\begin{array}{cc}
      0 & 0 \\
      0 & \sqrt{1-\gamma} \\
    \end{array}\right)$
&$\{\sqrt{\gamma}r_x,\sqrt{\gamma}r_y,r_z\}$ \\ \hline
\end{tabular}
}
\end{table}

The term damping channel typically encompasses two types of channels: \emph{amplitude damping} (AD) and \emph{phase damping} (PD). The energy dissipation effect that results from the loss of energy in a quantum system can be characterized by amplitude damping. On the other hand, phase damping refers to a noise process that results in a loss of quantum information but no energy loss. In the context of a single optical mode, the probability of losing a photon can be represented by $\gamma$. The operation elements and the changes in the state of the Bloch vector for these two damping channels are presented in Table \ref{Tab1}.

\subsection{Noisy GSA in Bloch}\label{NGSA}

A quantum state can be represented using a Bloch vector $\textbf{r}=(r_x,r_y,r_z)$. According to Equation (\ref{psi0}), the initial state $\ket{\psi_0}$ in GSA corresponds to the Bloch vector $r(0)=(\sin\theta,0,\cos\theta)$, where $r_y(0)=0$. Therefore, we will focus on the two dimensions of $r_x$ and $r_z$. The density matrix of $\ket{\psi_t}$ after $t$ iterations is given by:
\begin{align}
\rho(t)=G^t \rho(0){G^t}^\dag=\frac{1}{2}\left(
\begin{array}{cc}
1+r_z(t)&r_x(t)\\
r_x(t)&1-r_z(t)
\end{array}
\right).
\end{align}

The success probability can be expressed as:
\begin{eqnarray}\label{psk}
P(t)=\frac{1-r_z(t)}{2}.
\end{eqnarray}
As we can see, the success probability depends only on $r_z(t)$. To obtain an expression for $r_z(t)$, we will represent the Grover iteration in Bloch space.

The two operations $O$ and $R$, which act on a Bloch vector, can be represented in matrix form as in \cite{Rastegin}:
\begin{align}\label{ORB}
O_B=\left[
\begin{array}{cc}
-1&0\\
0&1
\end{array}
\right],
R_B=\left[
\begin{array}{cc}
-\cos2\theta&\sin2\theta\\
\sin2\theta&\cos2\theta
\end{array}
\right].
\end{align}
Hence, the ideal Grover iteration can be represented as:
\begin{align}\label{G}
G_B=\left[
\begin{array}{cc}
\cos2\theta&\sin2\theta\\
-\sin2\theta&\cos2\theta
\end{array}
\right].
\end{align}

In the presence of noise, the Grover iteration becomes $G_E$ in Bloch space, which will be discussed in Sections \ref{Sec3} and \ref{Sec4}. The value of $r_z(t)$ can then be calculated as follows:
\begin{align}
\left[
\begin{array}{cc} r_x(t)\\r_z(t) \end{array}
\right]
=G_E^t\left[
\begin{array}{cc} r_x(0)\\r_z(0) \end{array}
\right].
\end{align}

\section{GSA with diagonalizable noises in Bloch}\label{Sec3}

In this section, we introduce the concept of diagonalizable noise and examine the behavior of the GSA under such noises in the Bloch representation.

\newtheorem{def1}{Definition}\label{def1}
\begin{def1}[\textbf{Diagonalizable noise model}]

A matrix $M$ used to describe a noise model is said to be diagonalizable if it can be diagonalized in a representation.
\end{def1}

Since the initial state in GSA is represented by the Bloch vector $r(0)=(\sin\theta,0,\cos\theta)$ with $r_y(0) = 0$, we restrict our analysis to the two dimensions of $r_x$ and $r_z$.

\newtheorem{theorem1}{\textbf{Theorem}}
\begin{theorem}\label{Th1}

The Grover's search algorithm under the following noise channels, with the two dimensions of $r_x$ and $r_z$ in the Bloch representation, are diagonalizable: \textbf{Phase Flip}, \textbf{Bit Flip}, \textbf{Bit-Phase Flip}, and \textbf{Phase Damping}.
\end{theorem}

\begin{proof}
Consider the flip channels and let $\eta=1-2p$. In the Bloch representation of $r_x$ and $r_z$, the \emph{phase flip}, \emph{bit flip}, and \emph{bit-phase flip} can be represented as follows:
\begin{align}\label{Ef}
E_{p}=\left[
\begin{array}{cc}
\eta&0\\0&1
\end{array}
\right],\quad
E_{b}=\left[
\begin{array}{cc}
1&0\\0&\eta
\end{array}
\right]
,\quad
E_{bp}=\eta\left[
\begin{array}{cc}
1&0\\0&1
\end{array}
\right].
\end{align}
It is clear that these three types of noise are diagonalizable.

For the \emph{phase damping} noise, let $\sqrt{\gamma}=1-2p$. As noted in Table \ref{Tab1}, the actions of \emph{phase flip} and \emph{phase damping} are the same, and hence the noise is diagonalizable. This completes the proof.
\end{proof}


According to Definition 1, a noise model is considered non-diagonalizable if it cannot be diagonalized in a representation. For example, $AD$ is a non-diagonalizable noise in the two dimensions of $r_x$ and $r_z$ in the Bloch representation for Grover's search algorithm.  Now, we turn our attention to the properties of GSA with diagonalizable noises. The following theorem characterizes the effects of diagonalizable noises in the Grover iteration.  
\newtheorem{theorem2}{\textbf{Theorem}}
\begin{theorem}\label{Th2}
In the Bloch representation, the effects of diagonalizable noises in each reflection oracle are, except for the powers of the coefficients, equivalent to those in the whole Grover iteration.
\end{theorem}

\begin{proof}

Suppose that the diagonalizable noise is $E$, which contains two nonzero elements $e_1$ and $e_2$, represented as $E=diag\{e_1,e_2\}$.

First, we consider the effect of the diagonalizable noise on each reflection oracle, $G_E'=R_B\circ E\circ O_B \circ E$. Through some simple matrix calculations, we obtain:
\begin{align}\label{reoe}
G_E'&=R\circ E_B\circ O_B \circ E \nonumber\\
&=\left[
\begin{array}{cc}
\cos\theta e_1'^2&\sin\theta e_2'^2\\ -\sin\theta e_1'^2&\cos\theta e_2'^2
\end{array}
\right],
\end{align}
where $O_B$ and $R_B$ are given in Eq. (\ref{ORB}).

Next, we consider the effect of the noise on the entire Grover iteration, i.e., $G_E = R_B\circ E\circ O_B$ or $G_E = R_B\circ O_B\circ E$. Similar calculations give us:
\begin{align}\label{reo}
G_E&=R_B\circ E_B\circ O =R_B\circ O_B\circ E  \nonumber\\
&=\left[
\begin{array}{cc}
\cos\theta e_1&\sin\theta e_2\\ -\sin\theta e_1&\cos\theta e_2
\end{array}
\right].
\end{align}

If we set $e_1'^2 = e_1$ and $e_2'^2 = e_2$, then we have $G_E' = G_E$, as seen from the above two equations. This means that the effect of diagonalizable noise in each reflection oracle is the same as in the entire Grover iteration, except for the power of the coefficients.
\end{proof}

Importantly, the order of applying the noise $E$ and the Grover iteration $R_B\circ O_B$ matters, as they are not equal $R_B\circ O_B\circ E \neq E\circ R_B\circ O_B$. This means that the results of applying noise before or after the Grover iteration can lead to different outcomes.
However, Theorem \ref{Th2} may not be applicable for non-diagonalizable noises. In such cases, the analysis of noisy Grover search algorithm should consider applying the noise to each reflection oracle, as shown in previous works (Refs.\cite{Rastegin,Rastegin21}).
In the rest of this paper, we will focus on diagonalizable noises and the property stated in Theorem \ref{Th2} by considering the noise to act on the whole Grover iteration.

\section{GSA with diagonalizable noises in Bloch presentation}\label{Sec4}

In this section, we analyze the effect of various diagonalizable noises on the performance of the Grover search algorithm in the Bloch representation.

\subsection{Phase flip and phase damping noises in GSA}\label{SSecpf}

In this subsection, we will focus on the effects of phase flip and phase damping, two of the most significant noises in quantum information processing, on Grover's algorithm (GSA). There is no equivalent noise in classical information processing. We will use Theorems 1 and 2 to demonstrate that the impact of these two types of noise on GSA is similar.

GSA with phase damping noise in Bloch has been discussed by Rastegin \cite{Rastegin}. In that paper, the phase damping noise acted on the two oracles $O_B$ and $R_B$. For completeness and clarity, we briefly survey the main results. In this subsection, we will discuss how the phase flip noise acts on the whole Grover iteration $G$. In fact, the same results can be obtained for both of them according to Theorems \ref{Th1} and \ref{Th2}.

Rastegin's paper \cite{Rastegin} dealt with the impact of phase damping noise on the Bloch sphere representation of GSA. The phase damping noise  affects both the oracles $O_B$ and $R_B$. To summarize the key findings, this subsection will examine the impact of phase flip noise on the entire Grover iteration $G$. It can be noted that the same outcomes can be attained for both phase damping and phase flip noise, as demonstrated by Theorems 1 and 2.

In the two dimensions of $r_x$ and $r_z$, we have shown that for both phase flip and phase damping, $E_{p}=\begin{bmatrix} \eta & 0 \\0 & 1 \end{bmatrix}$. The noisy Grover iteration $G_E$ with phase flip is given by \begin{eqnarray}
G_{p}=R_B\circ O_B\circ E_{p}=\begin{bmatrix} \eta\cos 2\theta & \sin 2\theta \\ -\eta\sin 2\theta & \cos 2\theta \end{bmatrix},
\end{eqnarray} with eigenvalues \begin{eqnarray}
\lambda_\pm=A_+\pm iB,
\end{eqnarray} where \begin{eqnarray}\label{A}
A_\pm :=\frac{1\pm \eta}{2}\cos 2\theta
\end{eqnarray}
and \begin{eqnarray}\label{B}
B:=\left\{
\begin{aligned}
&\sqrt{\eta-A_+^2}, &&if\ \eta\geq A_+^2\\
&\sqrt{A_+^2-\eta}, &&if\ \eta<A_+^2.
\end{aligned}
\right.
\end{eqnarray}

According to Eqs. (\ref{A}) and (\ref{B}), we have $A_+^2 + B^2 = \eta$. Let $\phi$ be a positive angle such that
$\frac{A_+}{\sqrt\eta}= \cos\phi$, $\frac{B}{\sqrt\eta}= \sin\phi$ and
$\phi = \arctan\frac{B}{A_+}$.
The expressions allow us to simplify the eigenvalues. The resulting expressions are $\lambda_\pm=\sqrt{\eta}\exp(\pm i\phi)$. The eigenvectors can then be determined by solving $V^{-1}G_{p} V = D$, where $D$ is a diagonal matrix containing the eigenvalues $\lambda_\pm$ and $V$ is a matrix given by:
\begin{align}
V=\left[
\begin{array}{cc}
\sin 2\theta & \sin 2\theta\\
A_- + iB & A_- - iB
\end{array}
\right].
\end{align}

Since $G_{p}^t = VD^tV^{-1}$, the following expression for $G_{p}^t$ can be obtained:
\begin{align}
G_{p}^t=\frac{\eta^{t/2}}{B}\left[
\begin{array}{cc}
B\cos\phi t - A_-\sin\phi t & \sin\phi t \sin 2\theta\\
-\eta \sin\phi t \sin 2\theta & B \cos\phi t + A_-\sin\phi t
\end{array}
\right].
\end{align}

By considering that $r_x(0) = \sin \theta$ and $r_z(0) = \cos \theta$, the following expression for $r_z(t)$ is derived:
\begin{eqnarray}
r_z(t)  =\frac{\eta^{t/2}}{B} [-\eta\sin\phi t \sin 2\theta \sin\theta
+ (B \cos\phi t + A_- \sin\phi t) \cos \theta].
\end{eqnarray}
The success probability after $t$ iterations is given by:
\begin{eqnarray}\label{Pspf}
P_p(t)= \frac{1}{2}+\frac{\eta^{t/2}}{2B}
[\eta\sin\phi t\sin 2\theta \sin\theta
-(B\cos\phi t + A_- \sin\phi t)\cos\theta].
\end{eqnarray}

For ideal GSA, where $\eta=1$, we have $P_p(t)=P(t)=\sin^2\theta_t$. Thus, $P(t)$ is a periodic function that oscillates around $1/2$ and is bounded between 0 and 1. For $\eta<1$, the success probability tends towards $1/2$ quickly, since the second part of Eq. (\ref{Pspf}) decreases exponentially with increasing $t$.
In cases where GSA is searching in a large database with few targets, such that $m<<N$ and $\sin 2\theta \sin\theta$ can be ignored and $\cos\theta\simeq 1$, the success probability can be simplified as:
\begin{eqnarray}\label{Pspf2}
P_p(t)\simeq&&\frac{1}{2}-\frac{\eta^{t/2}}{2}(B\cos\phi t + A_- \sin\phi t)\nonumber\\
=&&\frac{1}{2}-\frac{\eta^{t/2}}{2B}r_p\sin(\phi t+\delta_p),
\end{eqnarray}
where $r_p=\sqrt{A_-^2+B^2}$ and $\delta_p=\arctan(B/A_-)$. Equation (\ref{Pspf2}) shows that the success probability $P_p(t)$ with phase noise is a periodic function of $t$ that oscillates around $1/2$ and decays exponentially towards $1/2$.

\subsection{Bit flip noise in GSA}\label{SSecbf}

In this subsection, we will focus on the effects of bit flip noise in GSA.
Theorem \ref{Th1} states that bit-flip noise can be expressed as $E_b=\left[
\begin{array}{cc}
1 & 0\\
0 & \eta
\end{array}
\right]$ in Bloch representation. As a result, the noisy Grover iteration, $G_E$, with bit-flip noise becomes:
\begin{align}
G_{b}=R_B\circ O_B\circ E_b
=\left[
\begin{array}{cc}
\cos 2\theta & \eta\sin 2\theta\\
-\sin 2\theta & \eta\cos 2\theta
\end{array}
\right].
\end{align}
Calculating the eigenvalues of $G_{b}$, we find that they are $\lambda_\pm=\sqrt{\eta}\exp({\pm i\phi})$, which are the same as the phase-flip. Further calculation of the corresponding eigenvectors gives $V^{-1}G_{b} V = D$, where $D = diag\{\lambda_+,\lambda_-\}$, \begin{align}
V=\left[
\begin{array}{cc}
- A_- -iB & -A_- +iB\\
\sin 2\theta & \sin 2\theta
\end{array}
\right]
\end{align}
and
\begin{align}
V^{-1}=\frac{1}{-2iB\sin{2\theta}}\left[
\begin{array}{cc}
\sin2\theta & A_- -iB\\
-\sin2\theta & - A_- +iB
\end{array}
\right].
\end{align}
Using $G_b^t = VD^tV^{-1}$, we get: 
\begin{align}
G_b^t=\frac{\eta^{t/2}}{B}\left[
\begin{array}{cc}
B\cos\phi t + A_-\sin\phi t & \eta\sin\phi t \sin 2\theta\\
-\sin\phi t \sin 2\theta & B \cos\phi t - A_-\sin\phi t
\end{array}
\right].
\end{align}
Therefore,
\begin{align}
r_z(t) =\frac{\eta^{t/2}}{B}[-\sin\phi t\sin 2\theta\sin\theta
+(B\cos\phi t-A_-\sin\phi t)\cos\theta],
\end{align}
and the success probability after $t$ iterations becomes:
\begin{align}
P_b(t) =\frac{1}{2}+\frac{\eta^{t/2}}{2B}[\sin\phi t\sin 2\theta\sin\theta
-(B\cos\phi t-A_-\sin\phi t)\cos\theta].
\end{align}

When the value of $\eta=1$, we have $P_b(t) = P(t)$, which represents the ideal scenario in GSA. However, when $\eta<1$, the success probability of GSA decreases exponentially with the increase in the number of iterations, settling around $1/2$, similar to the scenario with phase flip noise.
Using a similar method as for phase flip noise in \ref{SSecpf}, we analyze $P_b(t)$ and fingd the following expression:
\begin{eqnarray}\label{Psbf2}
P_b(t)\simeq\frac{1}{2}-\frac{\eta^{t/2}}{2B}r_b\sin(\phi t-\delta_b),
\end{eqnarray}
where $r_b=\sqrt{A_-^2+B^2}$ and $\delta_b=\arctan(B/A_-)$. Based on the above equation (\ref{Psbf2}), we can conclude that the success probability for GSA in the presence of bit flip noise is periodic, with an oscillating center around $1/2$, and it  decreases exponentially towards $1/2$.

\subsection{Bit-phase flip in GSA} \label{SSecbpf}

For bit-phase flip noise $E_{bp}$  in Bloch representation, we have:
\begin{align}
E_{bp}=\eta\left[
\begin{array}{cc}
1 & 0\\
0 & 1
\end{array}
\right]=\eta I,
\end{align}
where $\eta=1-2p$. As a result, the noisy Grover operator, $G_E$, becomes:
\begin{align}
G_{bp}=\eta\left[
\begin{array}{cc}
\cos 2\theta & \sin 2\theta\\
-\sin 2\theta & \cos 2\theta
\end{array}
\right]
=\eta G_B.
\end{align}
It is straightforward to obtain its eigenvalues $\lambda_\pm = \eta e^{\pm 2i\theta}$.
Therefore, the $r_z(t)$ becomes:
\begin{eqnarray}
r_z(t)=\eta^t\cos[(2t+1)\theta].
\end{eqnarray}

And the success probability can be calculated as:
\begin{eqnarray}\label{Pbp2}
P_{bp}(t)=\frac{1}{2}-\frac{\eta^t}{2}\cos[(2t+1)\theta].
\end{eqnarray}
As $\eta<1$ in a noisy environment, the success probability, $P_{bp}(t)$, decreases exponentially towards $1/2$ as $t$ increases and oscillates with a center around $1/2$, similar to the phase flip and bit flip noise cases.

We are particularly interested in the maximum success probability, $P_{max}$, and its corresponding iteration number, $t_m$. According to Eq. (\ref{Pbp2}), the bit-phase flip noisy Grover's algorithm will reach its maximum success probability in the first period. In the following part of this subsection, we will examine the special properties for $0 < t \leq t_m$.

\newtheorem{theorem3}{\textbf{Theorem}}
\begin{theorem}\label{Th3}

The effect of bit-phase noise on Grover's Search Algorithm (GSA) is both beneficial and detrimental. Initially, the success probability $p_{bp}(t)$ is slightly improved, but it then decreases exponentially. Bit-phase noise reduces the number of iterations $t_m$ required for the noisy Grover $G_{bp}$ to reach its maximum success probability $P_{max}$, but it also lowers the value of $P_{max}$.
\end{theorem}

\begin{proof}

The success probability after $t$ Grover iterations for the ideal GSA is given by:
\begin{eqnarray}\label{Pt}
P(t)=\sin^2\theta_t=\frac{1}{2}-\frac{1}{2}\cos[(2t+1)\theta].
\end{eqnarray}
Based on the properties of the cosine function and the fact that $\eta\leq1$, after comparing Equation (\ref{Pbp2}) with Equation (\ref{Pt}), we find that:
\begin{eqnarray}\label{Pbp3}
P_{bp}(t)\left\{
\begin{aligned}
&>P(t), &&if\ t< T/2\\
&=P(t), &&if\ t= T/2\\
&<P(t), &&if\ T/2<t\leq t_m.
\end{aligned}
\right.
\end{eqnarray}

This means that when the bit-phase flip noise is present in the noisy GSA, its success probability is better than in the ideal case for the time period $t< T/2$. As a result, the presence of bit-phase flip noise increases the success probability initially, but it reduces the maximum success probability eventually.

According to Eqs. (\ref{Pbp2}) and (\ref{Pbp3}), it is established that $P_{bp}(T/2)=P(T/2)=1/2$ and $P_{bp}(t)\geq 1/2$ for $t\geq T/2$. In simpler terms, regardless of the intensity of the bit-phase flip noise, GSA requires only $O(\sqrt{N/m}/2)$ iterations to achieve a success probability of no less than 1/2. As a result, even in the presence of bit-phase flip noise, GSA remains a superior option compared to the classical stochastic algorithm with an average number of queries being $N/2$ and a success probability of $1/2$. However, if a success probability greater than 1/2 is desired, the bit-phase flip noise will always have a detrimental impact on the performance of GSA as $P_{bp}(t)<P(t)$ for $t>T/2$ as indicated by Eq. (\ref{Pbp3}).

To determine $t_m$ and $P_{max}$, we calculate the derivative of $P_{bp}(t)$ with respect to $t$. As we require $P_m \geq 1/2$, we find that $\theta_{t_m}/2 \in [\pi/4,\pi/2]$, which translates to $T/2 \leq t_m \leq T$, as indicated by Eq. (\ref{Pbp3}). The following results are obtained:
\begin{eqnarray}\label{tmax}
t_m=\left\lfloor \frac{\arctan(\ln\eta/2\theta) + \pi}{2\theta} \right\rfloor,
\end{eqnarray}
and
\begin{eqnarray}\label{Pmax}
P_{max}=\frac{1}{2}-\frac{\eta^{t_m}\cos(2\theta t_m + 1)}{2}.
\end{eqnarray}
With the help of Eqs. (\ref{tmax}) and (\ref{Pmax}), we can calculate the maximum success probability for a given $\eta$. For $\eta=1$, we get $t_m = T \simeq \pi/4\sqrt{N/m}$ and $P_{max} \simeq 1$ when $m \leq N$, which corresponds to the ideal GSA scenario.

Based on Eqs. (\ref{tmax}) and (\ref{Pmax}), both $t_m$ and $P_{max}$ decrease as the value of $\eta$ decreases (which means more noise is in the system) for $0<\eta\leq 1$ when $t<t_m$. 
In other words, bit-phase noise reduces the number of iterations $t_m$ required for the noisy Grover $G_{bp}$ to reach its maximum success probability $P_{max}$, but it also lowers the value of $P_{max}$.
\end{proof}

According to Theorem \ref{Th3}, the impact of bit-phase flip noise on the GSA is two-fold. On one hand, the GSA requires fewer iterations to reach its maximum success probability under noisy conditions. On the other hand, as $\eta$ decreases, the maximum success probability $P_{max}$ decreases. This implies that while noise may help reduce the number of iterations required, it negatively impacts the performance of the algorithm in terms of success probability.

\section{Numerical results about GSA with diagonalizable noises}\label{Sec5}

In this section, we present numerical results to illustrate the performance of GSA in the presence of different diagonalizable noises.

Consider a database with $N=2^8$ items and the task of searching for a single target. We present the success probability of GSA as a function of the number of search iterations, $t$, while varying the noise level from $\eta=1$ to $\eta=0.7$. The noise level of $\eta=1$ corresponds to the ideal GSA without any noise. To clearly show the change in the maximum success probabilities under different noise levels, we mark them with star markers in our results.

\begin{figure}[H]
\centering{\includegraphics[width=0.55\textwidth]{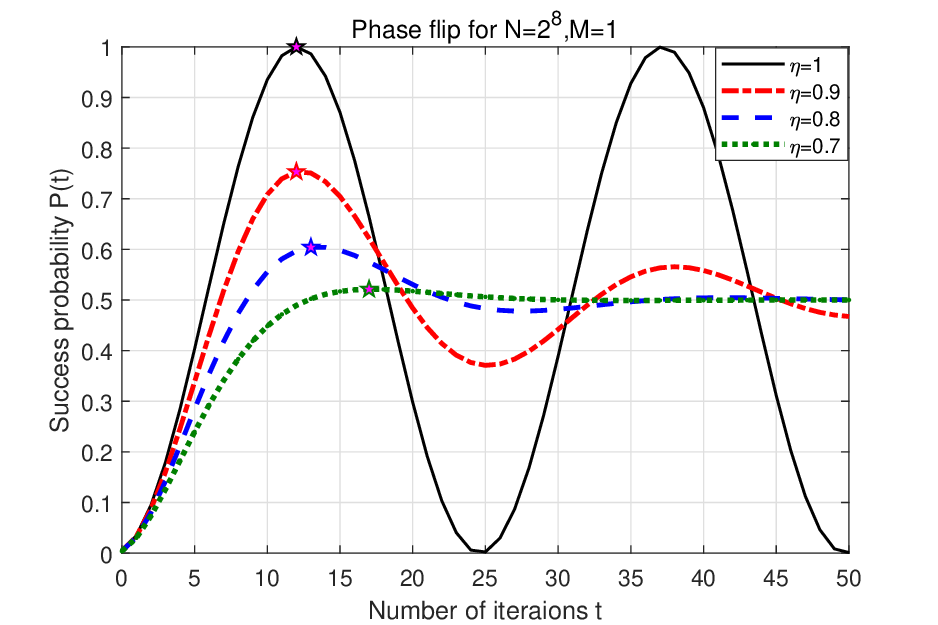}}
\caption{\label{fig:pf}(Color online) Success probability vs. search iteration steps for Phase flip noise.}
\end{figure}

\begin{figure}[H]
\centering{\includegraphics[width=0.55\textwidth]{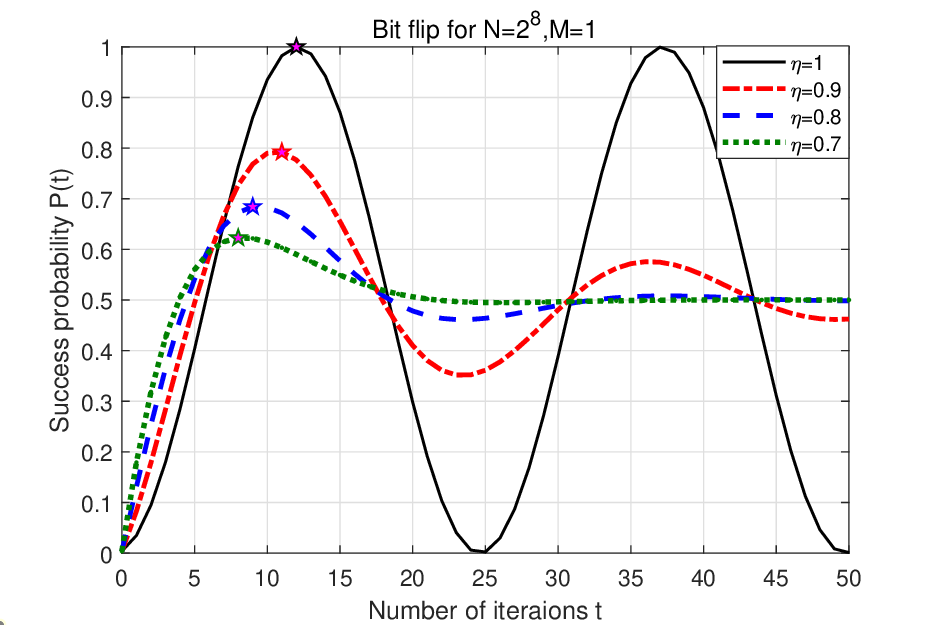}}
\caption{\label{fig:bf}(Color online) Success probability vs. search iteration steps for Bit flip noise.}
\end{figure}

\begin{figure}[H]
\centering{\includegraphics[width=0.55\textwidth]{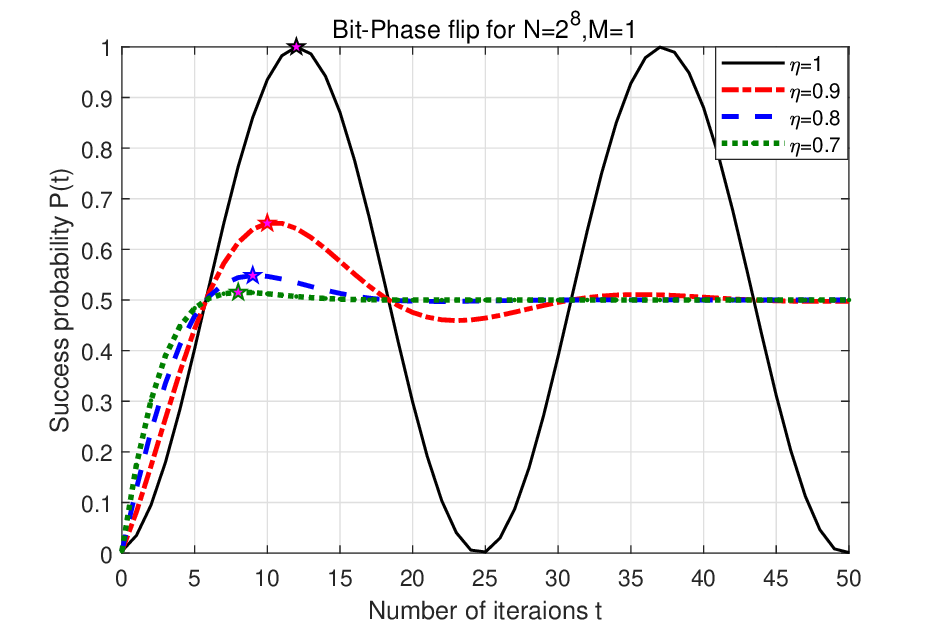}}
\caption{\label{fig:bpf}(Color online) Success probability vs. search iteration steps for Bit-Phase flip noise.}
\end{figure}

The performance of GSA with different types of noise is demonstrated in Figs. \ref{fig:pf}, \ref{fig:bf}, and \ref{fig:bpf} for phase flip, bit flip, and bit-phase flip noises, respectively. These figures illustrate the dynamic process of the algorithm in terms of the success probability as a function of the number of search iterations $t$.

As shown in the figures, the success probabilities oscillate around $1/2$ both in the absence and presence of noise. With increasing noise levels, the success probabilities not only decay more quickly but also exhibit lower maximum probabilities, $P_{max}$. This trend continues until the success probabilities approach $1/2$ as the number of iterations, $t$, increases, as previously documented in literature \cite{Rastegin}.

For phase flip noise, Fig. \ref{fig:pf} shows that the number of iterations needed to reach $P_{max}$ increases as the noise level increases. If the noise is too strong, this number may become too large, resulting in the loss of the quantum advantage of the GSA.

However, when the GSA is affected by bit flip or bit-phase flip noises, the situation is different, as depicted in Figs. \ref{fig:bf} and \ref{fig:bpf}. Despite a decrease in $P_{max}$, the number of iterations required to reach $P_{max}$, $t_m$, is reduced compared to the ideal GSA with $\eta=1$.
Initially, the success probabilities are improved and increase with the noise levels for $t<T/2$. Furthermore, when $T/2\leq t<T$, the success probability $P$ remains greater than or equal to $1/2$, regardless of the magnitude of the noise. This means that the GSA, even under the influence of bit flip or bit-phase flip noise, always requires only $O(\sqrt{N/m}/2)$ iterations to reach a success probability of at least $1/2$, which is superior to the classical stochastic algorithm that requires $N/2$ queries on average with a success probability of $1/2$.

Let us take a closer look at the maximum probabilities, which are marked with stars in the figures. It becomes evident that, for the same level of noise, the maximum success probabilities for GSA under bit-phase flip noises are the lowest. This indicates that GSA is the most vulnerable to bit-phase flip noise among the four types of noise. On the other hand, phase flip and phase damping noises are the most detrimental to GSA as they cause the success probabilities to decline as the number of iterations $t$ increases. These noises can completely erode the quantum advantage of GSA.

\section{Discussion}\label{Sec6}

The oracle-based model is a prevalent approach in quantum algorithms, where the focus is on the number of queries made, rather than the internal structure of the oracle. As the exact internal structure of the oracle is often unknown, it becomes challenging to assess the impact of noise on its performance. In this study, we adopt a simplified approach by considering the oracle to collectively suffer from different types of noise, as previously explored by Rastegin in \cite{Rastegin}. The performance of GSA under various types of noise, including phase flip, phase damping, bit flip, and bit-phase flip, were studied in this research.

(1) In the theoretical aspect, we analyzed the impact of noise on the search performance of GSA. In Sec. \ref{Sec4}, we established the relationship between the success probability, noise level, and number of search iterations. Our findings indicate that the success probability of GSA decreases exponentially with a periodic and oscillating center around $1/2$ as the number of iterations increases. Eventually, when the number of iterations becomes large enough, the success probability converges to $1/2$.
More remarkably, we discovered some novel and intriguing results. By thoroughly examining the maximum probabilities and the corresponding iteration numbers for GSA with bit-phase flip noise, we found that they can reach a success probability of $1/2$ with fewer than $O(\sqrt{N/m})$ iterations. In comparison to classical search algorithms, which typically require an average of $N/2$ iterations to achieve the same success probability, the noisy GSA outperforms them in terms of iteration efficiency.

(2) The numerical experiments conducted in this study provide us with some unique insights into the effect of different noise levels on the performance of the noisy GSAs. In the case of phase flip and phase damping noise, as long as the noise levels are not excessive, we observe that the maximal success probability consistently decreases until it reaches $1/2$. As the noise levels increase, the performance of the noisy GSA will deteriorate, and it will require a larger number of iterations to reach the success probability of $1/2$.
On the other hand, for bit flip and bit-phase flip noise, we have made a surprising discovery: in the initial stage of the search, these noises can improve the success probability, but as the number of iterations increases, the maximal success probability will eventually reduce to $1/2$. Furthermore, these noises can help the noisy GSA reach the maximal success probability of $1/2$ with fewer iterations.
While it may seem that noise can help the GSA improve its performance in some aspects, it's important to note that the success probability reaching $1/2$ could also be due to the noise being so strong that the oracles can no longer distinguish between the target and nontarget states, leading to an inability to find the right target state.
It's worth mentioning that in this study, we consider the noises to be acting collectively on the oracles, and the states are represented by Bloch vectors in the two dimensions of $r_x$ and $r_z$. Moreover, the bit flip noise flips the state $\ket{\chi_1}$ to $\ket{\chi_0}$ instead of flipping $\ket{1}$ to $\ket{0}$ or vice versa, as the space is spanned by the target state $\ket{\chi_1}$ and the nontarget state $\ket{\chi_0}$.

(3) Comparison with previous research: Our results are consistent with most previous research in that noise has a negative impact on the performance of GSA. For small noise levels, the success probability decreases periodically. Our results are similar to the studies in literature such as \cite{Rastegin,Rastegin21} that discuss phase damping and amplitude damping noises, which cause the success probability to exponentially decrease and eventually tend to $1/2$. However, our results differ from some previous studies, such as \cite{Cohn,Reitzner}, which consider local noises that decrease the success probability to a small value as the noise level exceeds a specific threshold.
Different types of noise can highlight different aspects of the impact of noise on GSA, and it is challenging to fully characterize noise in real-world NISQ devices \cite{Christopher2020}. Collective noise, as studied in our research, can also shed light on some characteristics of noisy GSA.

\section{Conclusion}\label{Sec7}

In the NISQ era, noise can have a negative impact on the performance of Grover's search algorithm (GSA), potentially rendering it ineffective. Our analysis and simulations revealed that for various typical diagonalizable collective noises, the success probability of GSA decays towards $1/2$ with an oscillation centered around $1/2$. Although noise in general can hinder the performance of GSA, there are exceptions. Certain types of noise, such as  bit flip and bit-phase noises, may actually enhance the performance of GSA in specific stages of the search process.
This finding presents a potential approach to managing noise in NISQ computing, by utilizing it instead of simply trying to eliminate it. Based on this principle, we propose the development of quantum algorithms that take advantage of noise, as previously discussed in literature \cite{Du2021,Aida2021}. With moderate amounts of noise, we can confirm the accuracy of quantum algorithms on NISQ devices and apply noise in areas like machine learning, which has been investigated  in the NISQ era.

\begin{acknowledgements}
We thank the anonymous reviewers and editors for their valuable comments and suggestions which have been a great help for us to improve the quality of this paper.
This work was supported by the National Natural Science Foundation of China (Grant No. 61772565), Guangxi Science and Technology Program (No. Guike AD21075020), Guangxi Natural Science Foundation (No. 2019GXNSFBA245087), Guangxi Key Laboratory of Cryptography and Information Security (No.202133), the Guangdong Basic and Applied Basic Research Foundation (Grant No. 2020B1515020050), the Key Research and Development Project of Guangdong Province (Grant No. 2018B030325001) and the Major Key Project of PCL, Innovation Program for Quantum Science and Technology (2021ZD0302900).

\end{acknowledgements}

\section*{Declarations}

{\bf Data availability statement:} No data were used during the study. All code that supports the findings of this study are available from the corresponding author and Minghua Pan upon reasonable request.

{\bf Conflicts of Interest:} The authors declare no conflict of interest.


\bibliography{RefQIP2022}

\end{document}